\documentclass[]{aa}

\usepackage{graphicx}
\usepackage{adjustbox}
\usepackage{txfonts}
\usepackage{url}
\usepackage{relsize}
\usepackage{color}

\begin{document}

\def\simlt{\mathrel{\rlap{\lower 3pt\hbox{$\sim$}}\raise 2.0pt\hbox{$<$}}}
\def\simgt{\mathrel{\rlap{\lower 3pt\hbox{$\sim$}} \raise 2.0pt\hbox{$>$}}}

   \title{Average fractional polarization of extragalactic sources \\ at \textit{Planck} frequencies }

    \author{T. Trombetti\inst{1,2,3}\fnmsep\thanks{e-mail:trombetti@ira.inaf.it}
                     \and
           C. Burigana\inst{1,2,4}\fnmsep\thanks{e-mail:burigana@ira.inaf.it}
           \and
           G. De Zotti\inst{5}\fnmsep\thanks{e-mail: gianfranco.dezotti@oapd.inaf.it}
           \and
           V. Galluzzi\inst{6}\fnmsep\thanks{e-mail: vincenzo.galluzzi@inaf.it}
           \and
           M. Massardi\inst{1}\fnmsep\thanks{e-mail: massardi@ira.inaf.it}
           }

    \institute{INAF, Istituto di Radioastronomia, Via Piero Gobetti 101, I-40129 Bologna, Italy
    \and
    Dipartimento di Fisica e Scienze della Terra, Universit\`a di Ferrara, Via Giuseppe Saragat 1, I-44122 Ferrara, Italy
    \and
    INFN, Sezione di Ferrara, Via Giuseppe Saragat 1, I-44122 Ferrara, Italy
    \and
    INFN, Sezione di Bologna, Via Irnerio 46, I-40127 Bologna, Italy
    \and
    INAF, Osservatorio Astronomico di Padova, Vicolo dell'Osservatorio 5,
I-35122 Padova, Italy
    \and
    INAF, Osservatorio Astronomico di Trieste, Via Giambattista Tiepolo 11, I-34143 Trieste, Italy
            }

   \date{Received ...; accepted ...}

  \abstract{Recent detailed simulations have shown that an insufficiently accurate
  characterization of the contamination of unresolved polarized extragalactic sources
  can seriously bias measurements of the primordial cosmic microwave
background (CMB) power spectrum if the tensor-to-scalar
  ratio $r\sim 0.001,$ as predicted by models currently of special interest (e.g., Starobinsky's $R^2$ and Higgs inflation).
  This has motivated a reanalysis of the median polarization fraction of extragalactic sources
  (radio-loud AGNs and dusty galaxies) using data from the \textit{Planck} polarization maps.
  Our approach, exploiting the intensity distribution analysis, mitigates or overcomes the most
  delicate aspects of earlier analyses based on stacking techniques. By means of simulations,
  we have shown that the residual noise bias on the median polarization fraction,
  $\Pi_{\rm median}$, of extragalactic sources is generally $\simlt 0.1\%$.
For radio sources, we have found $\Pi_{\rm median} \simeq 2.83\%$, with no
significant dependence on either frequency or flux density, in good agreement
with the earlier estimate and with high-sensitivity measurements in the
frequency range 5--40\,GHz. No polarization signal is detected in the case of
dusty galaxies, implying 90\% confidence upper limits of $\Pi_{\rm dusty}\simlt
2.2\%$ at 353\,GHz and of $\simlt 3.9\%$ at 217\,GHz. The contamination of CMB
polarization maps by unresolved point sources is discussed. }


   \maketitle


\section{Introduction}

The detection of the primordial B-mode polarization of the  cosmic microwave
background (CMB) is currently the most pressing question of particle physics and cosmological
research because the signal carries a clean signature of primordial
inflation. It is very faint, however, because it is generated by tensor perturbations,
which are so weak as to defy detection so far. The tightest upper limit on the
ratio, $r$, of tensor-to-scalar fluctuations at the pivot scale of
$0.05\,\hbox{Mpc}^{-1}$ is $r<0.07$ at the 95\% confidence level
\citep{BICEP2016}.

Measurements of Galactic \citep{ChoiPage2015,
PlanckCollaboration_pol_dust2016,  PlanckCollaboration_low_freq_fore2016,
Krachmalnicoff2016, PlanckCollaboration_pol_dust2017} and estimates of
extragalactic \citep{Tucci2004, TucciToffolatti2012, Bonavera2017radio,
Bonavera2017dusty} polarized foreground emission show that even in the
cleanest 70\% of the sky, this emission {}dominates both primordial B-modes and
instrumental noise of the most sensitive forthcoming or proposed CMB
polarization experiments, such as
the Lite (Light) satellite for the studies of B-mode polarization and Inflation from cosmic background Radiation Detection (LiteBIRD),\footnote{\url{http://litebird.jp/}}  
the Cosmic ORigins Explorer (CORE) \citep{Delabrouille2018}, 
the Probe of Inflation and Cosmic Origin (PICO), one of the eight Probe-Scale space missions in the range of \$\,400M -- \$\,1000\,M
that are being funded by NASA,\footnote{\url{https://zzz.physics.umn.edu/groups/ipsig/cmbprobe2016proposal}}
the Primordial InflaXIon (Inflation) Explorer (PIXIE) \citep{Kogut2016} 
and the CMB-Stage IV experiment (CMB-S4) \citep{Abazajian2016}.

Thus, the capability of future CMB observations to measure or constrain CMB
B-mode power spectra will be limited not by sensitivity of the instrument, but
by the ability of removing foreground contamination with extreme accuracy.
According to \citet{Delabrouille2018}, foreground cleaning must reach at least the 99.9\% level at $\ell \simeq 10$,
at least the 99\% level at $\ell \simeq 100$ and at least the 90\% level at
$\ell \simeq 1000$ to bring the residual
contamination
below noise and/or cosmic variance in multipole bins $\Delta\ell/\ell=
0.3$.

Furthermore, the B-mode power spectrum exceeds that of primordial origin for $\ell \simgt 10$ if  $r\simlt 10^{-2}$ due to gravitational lensing \citep[cf.
Fig.~1 of][]{Delabrouille2018}. Hence the detection of primordial B modes
requires a very accurate control of lensing effects, which in turn are
contaminated by fluctuations of unresolved extragalactic sources.

There is no definitive prediction for the magnitude of the tensor-to-scalar
ratio,  $r$. For inflationary models driven by a fundamental scalar field, the
value of $r$ is related to the total field excursion \citep[i.e., to the
inflation field range in Planck units;][]{Lyth1997, BaumannMcAllister2007}.
Models currently of special interest (e.g., Starobinsky's $R^2$ and Higgs
inflation) predict $r \sim 0.003$  \citep{Martin2014, Abazajian2016,
Finelli2016}. Large-field  inflation with super-Planckian field excursions
implies even lower values, $r\simlt 0.001$.

Accurate simulations \citep{Remazeilles2018} showed that for $r\simeq
10^{-3}$, the overall uncertainty on $r$ is dominated by foreground residuals
and that unresolved polarized point sources can be the dominant foreground
contaminant over a broad range of angular scales ($\ell \simgt 50$). An
accurate understanding of the polarization properties of extragalactic sources
is therefore crucial.

Root-mean-square temperature fluctuations due to unresolved extragalactic
sources have a minimum around 120 GHz \citep[cf. Fig.~2 of][]{DeZotti2015},
corresponding to the transition from the dominance of radio sources to that of
dusty galaxies. The main radio source populations in the frequency range of CMB
experiments (the so-called `cosmological window') are  the compact flat- and
inverted-spectrum populations, primarily blazars (BL Lac objects and flat-spectrum
radio quasars).

Our understanding of polarization properties of extragalactic radio sources  in
the cosmological window is still quite poor \citep[for reviews,
see][]{TucciToffolatti2012, GalluzziMassardi2016}.

Direct WMAP detections in polarization are limited to small numbers of sources
\citep{Wright2009, LopezCaniego2009}. \textit{Planck}, thanks to its higher
sensitivity and angular resolution, has yielded more detections. The Second
\textit{Planck} Catalog of Compact Sources lists over 120 objects with
polarized emission significant at a $> 99.99\%$ level, not considering the
PCCS2E sub-catalog, whose reliability is unknown \citep[Table~14
of][]{PCCS2_2016}. These sources, however, mostly lie at low Galactic latitudes
so that only a minor fraction of them are expected to be extragalactic. The
number of detections in the extragalactic zone ($|b| > 30^\circ$) ranges
from 28 at 30\,GHz to $\sim 10$ in the channels up to 217\,GHz, to 1 at
353\,GHz \citep{DeZotti2018}; the two highest \textit{Planck} frequencies (545
and 857\,GHz) were not polarization sensitive. All detected sources are
radio-loud AGNs.

Follow-up polarization measurements at 8.4, 22, and 43 GHz of a complete sample
of 199 extragalactic sources stronger than 1 Jy in the 5 yr WMAP catalog were
carried out by \citet{Jackson2010}. Polarimetric observations of 211 radio-loud
active galactic nuclei (AGNs) at 86 and 229\,GHz were performed by
\citet{Agudo2014}. Their $\ge 3\,\sigma$ detection rate was 88\% at 86\,GHz
and 13\% at 229\,GHz. The sample selection was designed to be flux limited
at 1 Jy at 86\,GHz; however, 51\% of the sources were found to have $S_{86}< 1\,$Jy
and 22\% even had $S_{86}< 0.5\,$Jy, probably due to variability.

Most recently, \citet{Galluzzi2017a, Galluzzi2018} presented high-sensitivity
($\sigma_P \simeq 0.6\,$mJy) polarimetric observations at seven frequencies, from
2.1 to 38\,GHz, of a complete sample of $104$ compact extragalactic radio
sources brighter than $200\,$mJy at $20\,$GHz. The observations achieved a
$5\sigma$ detection rate of 90\%. Individual sources showed a broad variety of
spectral shapes (flat, steep, upturning, peaked, inverted, downturning) both in
total intensity and in polarization, but with substantial variations with
frequency of the polarization fraction from one object to another. Because of this
complexity, extrapolations to frequencies of CMB experiments to mitigate the
point source contamination of CMB maps are unreliable.

In the case of star-forming galaxies, the polarized emission above 100\,GHz is
dominated by dust.
At lower frequencies, the synchrotron emission takes over,
but at these frequencies, the extragalactic sky is
far from being dominated by radio sources.
Polarization properties of dusty galaxies as a whole at (sub-)millimeter
wavelengths are almost completely unexplored. The only published measurement
\citep{GreavesHolland2002} yielded a polarization fraction $\Pi = 0.4\%$ for
the prototype starburst galaxy M\,82. Integrating the \textit{Planck} dust
polarization maps, \citet{DeZotti2018} found an average value of the Stokes $Q$
parameter of about 2.7\%. If this is typical for late-type galaxies seen
edge-on and the polarization fraction scales as $\cos(\theta)$, $\theta$ being
the inclination angle, the mean polarization fraction, averaged over all
possible inclinations, should be $\simeq 1.4\%$.

It is particularly important to characterize the point source contamination in
the 60--120\,GHz frequency range, where \textit{Planck} data
\citep{PlanckForegroundMaps2016} have shown that the brightness temperature
spectra of diffuse polarized foregrounds display a broad minimum \citep[cf.
Fig. 1 of][]{Remazeilles2018}.

Although the number of sources detected in polarization by \textit{Planck}  in
the extragalactic zone is quite limited, estimates of the mean polarization
fraction of fainter sources can be obtained using stacking techniques, that is, by
co-adding the polarized signal from many objects detected in total intensity
but not in polarization, to increase the signal-to-noise ratio (S/N). A first
attempt in this direction was carried out by \citet{Bonavera2017radio} for
radio sources and by \citet{Bonavera2017dusty} for dusty galaxies.

\citet{Bonavera2017radio} applied the stacking techniques to the 1560 30\,GHz
sources in the Second \textit{Planck} Catalog of Compact Sources
\cite[PCCS\,2;][]{PCCS2_2016}, spanning about a factor of 20 in flux density,
and followed them in all \textit{Planck} maps in polarization (at 30, 44, 70,
100, 143, 217, and 353 GHz). The subsample outside the \textit{Planck} GAL60
mask (covering about 40\% of the sky, around the Galactic plane) and outside
the Magellanic Cloud regions, contains 881, likely extragalactic, radio
sources. The remaining 679 are probably mostly Galactic.

The application of stacking to high-resolution data is relatively
straightforward,  since only two ingredients need to be included: the (faint)
signal of sources, and the noise. By coadding measurements at the positions of
$n$ sources, the signals add up while Gaussian noise decreases as $n^{-1/2}$.
In the case of the low-resolution \textit{Planck} data, the situation is much
more complicated because each resolution element containing a source also
contains other polarized signals: the CMB itself, and diffuse synchrotron and
dust emissions from the Galaxy. The Galactic contamination is particularly
difficult to include because of its highly non-Gaussian statistics and its
significant variations on the sky.

Furthermore, the mean polarized flux density, $P$, is computed from the Stokes
parameters $Q$ and $U$ as $P=(Q^2+U^2)^{1/2}$. Any quadratic sum of this
kind is liable to the so-called noise bias, however
\citep[e.\,g.,][]{WardleKronberg1974}: the errors on $Q$ and $U$ add a
contribution to $P$. The standard methods for correcting for this bias cannot be
applied to stacking because sources are not individually detected.

Again, because sources are not detected, the desired result, that is, the mean
polarization fraction, $\langle\Pi\rangle= \langle P/S \rangle,$ where $S$ is
the total flux density, cannot be computed directly but is approximated by
$\langle P \rangle/\langle S \rangle$. Recent studies (at lower frequencies) did
not find evidence of systematic variations of the \textit{\textup{mean}} polarization
fraction with $S$ \citep{Hales2014, Galluzzi2017a}, but on the other hand, the
spectra of polarized emission of individual sources are generally substantially
different from those in total intensity \citep{Galluzzi2017a, Galluzzi2018}.
In addition, for the faintest sources the polarized signal is dominated by other
components or by noise, so that it is unrelated to $S$. The combination of
these effects may be a quite delicate point, in particular for flux
densities differing by more than one order of magnitude. As described in next
section, our approach does not rely at all on the $\langle\Pi\rangle= \langle
P/S \rangle \simeq \langle P \rangle/\langle S \rangle$ approximation.

When applied in this context, stacking techniques typically require simulations
to include the noise bias. For example, the corresponding correction found in
\citet{Bonavera2017radio} is lower than $\simeq 20$\% when applied to
$\sqrt{\langle\Pi^2\rangle,}$ but reaches a factor of  approximately 3 to 6,
depending on frequency, when applied to $\langle\Pi\rangle$.

\citet{Bonavera2017dusty} applied the same approach to a sample of 4697 dusty
galaxies  drawn from the PCCS2 857\,GHz catalog, deriving average corrected
polarization fractions and corresponding median values at 217 and 353\,GHz,
with a tentative detection at 143\,GHz.

Given the real, substantial difficulty of the problem and the importance of
reaching an assessment as solid as possible of the mean polarization properties
of extragalactic sources in the \textit{Planck} frequency range, we decided to
carry out a new investigation adopting an independent approach. In this paper
we present a simpler analytical approach and describe how we control the
critical aspects.

The layout of the paper is the following. In Sect.~\ref{sect:method} we
outline our method. In Sect.~\ref{sect:validation} we validate it by applying
it to sources whose polarization has been measured by \citet{Jackson2010} at
43\,GHz and by \citet{Agudo2014} at 86 and 229\,GHz. In
Sect.~\ref{sect:results} we report our results, which are summarized in
Sect.~\ref{sect:conclusions}.

\section{Outline of the method}\label{sect:method}

Our approach uses the intensity distribution analysis
\citep[IDA;][]{DeZotti1989, Barcons1995}.  Briefly, this method consists  of
measurements of signals in a map at the positions of a given source catalog.
The distribution of signals is compared with that for the blank sky, measured
at random positions, away from sources (control fields). If some statistical
test  detects a significant difference, in the sense that the source
distribution is shifted toward higher values than that of control fields, a
signal is detected. For this purpose, we use the one-sided Kolmogorov-Smirnov
(KS) statistics.

The polarized flux density of sources is then estimated as
\begin{equation}
P = \left(P^2_{\rm s} - P^2_{\rm CF, median}\right)^{1/2},
\label{eq:pol_flux}
\end{equation}
\noindent where $P_{\rm s}$ is the polarized signal in \textit{Planck} maps at
the source position  and $P^2_{\rm CF, median}$ is the median value of $P^2$ in
the control fields. The subtraction removes, in a statistical sense, the
contributions to $P$ of the noise, of the CMB, and of polarized Galactic
emissions. It thus largely corrects for the noise bias, as verified
through a comparison with direct polarimetric measurements
(Sect.~\ref{sect:validation}) and via simulations (Sect. \ref{sect:sims}).
When $P^2_{\rm s} < P^2_{\rm CF, median}$ , we set
$P=0$.\footnote{Since we computed median values, which requires positive
signals for more than 50\% of the sources, this choice does not affect our results.}

We have considered radio sources listed in the Second \textit{Planck} Catalog
of  Compact Sources \citep[PCCS2;][]{PCCS2_2016}, selecting those detected at
143\,GHz and located at high Galactic latitude ($|b|\ge 20^\circ$),  excluding
the areas inside the mask adopted for the \textit{Planck} polarization analysis
at 100\,GHz at high resolution
(COM\_Mask\_Likelihood-polarization-100\_2048\_R2.00.fits). We further excluded
sources flagged as extended in the PCCS2. We repeated the analysis with a
Galactic cut at $|b|=30^{\circ}$ , obtaining consistent results but with a poorer
statistics.  Only the results for the cut at $|b|=20^{\circ}$ are reported.

At high Galactic latitudes, objects above the the PCCS2 detection limit at
143\,GHz  are almost exclusively radio sources \citep[cf. ][]{Negrello2013,
PlanckCollaboration2013stat_prop, Mocanu2013}. The contaminating fraction of
dusty galaxies or Galactic sources is negligibly small at low frequencies, but
increases with increasing frequency because dust emission steeply rises with
frequency at millimeter/submillimeter wavelengths. An analysis of the brightest flux density
bins at 353\,GHz ($S_{353}> 1780\,$mJy) has shown that about half of sources
are either Galactic HII regions or nearby dusty galaxies. These objects are
characterized by a strong upturn of the spectrum at submilimeter wavelengths that is due to
dust emission. Thus, their contamination of the lower frequency samples is
negligibly small. Nevertheless, they were excluded from all samples.

The \textit{Planck} light maps in temperature ($T$) and in polarization
(Stokes  parameters $Q$ and $U$) with $N_{\rm side} = 1024$ at 30 and 44 GHz,
and $N_{\rm side} = 2048$ at higher frequencies were inspected at the source
positions. For each source we summed the (positive and negative) signals in all
pixels within a circle with FWHM/2 radius. Temperature and polarization
signals, given in temperature units, were converted into flux densities using the
unit conversion software provided by the \textit{Planck}
Collaboration\footnote{\url{http://wiki.cosmos.esa.int/planckpla2015/index.php/Unit_conversion_and_Color_correction}}
\citep{PlanckCollaborationV2014, PlanckCollaborationIX2014}. We have checked
that we obtain good agreement with the PCCS2 flux densities in
this way.

The procedure was repeated for a large number of control fields. The
centers of  these fields were required to be separated by at least 2\,FWHM from
each other and from the sources. This was efficiently implemented using the
properties of the hierarchical equal area isolatitude pixelation
\citep[HEALPix;][]{Gorski2005}.\footnote{\url{http://healpix.sourceforge.net}}
It was enough to place the control field centers at the centers of the $N_{\rm
side} = 64$ and of the $N_{\rm side} = 128$ pixels for the frequencies of the
Low Frequency Instrument (LFI; 30, 44 and 70 GHz) and of the High Frequency
Instrument (HFI; 100, 143, 217 and 353\,GHz), respectively. We selected
26,231 and 111,442 control fields at the LFI and HFI frequencies, respectively.

Unlike the stacking technique, the IDA considers each source
individually,  therefore it allows us to compute the mean and the median $P/S$
ratios directly. We preferred the median to the mean values because the latter are
less stable, being very sensitive to the presence of a few objects with
exceptionally high polarized flux densities. Analogously to the polarized flux
density, the total flux density of sources is estimated by subtracting, in a
statistical sense, the other contributions (mainly due to Galactic emissions)
as
\begin{equation}
S = S_{\rm s} - S_{\rm CF, median} \,,
\label{eq:tot_flux}
\end{equation}
\noindent where $S_{\rm s}$ is the signal in \textit{Planck} maps at the source
position and $S_{\rm CF, median}$  is the median value of $S$ in control
fields. Thus, the source polarization fraction is estimated by
\begin{equation}
\Pi = P/S \,
\label{eq:poldeg}
,\end{equation}
\noindent
with $P$ and $S$ given by Eqs. (\ref{eq:pol_flux}) and (\ref{eq:tot_flux}), respectively.

In order to investigate the possible flux-density dependence of the median
polarization degree  on flux density, we subdivided the samples selected at
each frequency into flux density bins containing 30 sources each, starting from
the 90\% completeness limit given by the PCCS2 paper. This left fewer than 30
objects in the brightest bin, which has the strongest signal, which is also detectable
with fewer sources. Since the dusty sources were removed after the bins were defined, the final number of sources in some bins is
also slightly smaller than $30$.  For the source binning we used the
DETFLUX photometry, as recommended for point sources up to 217\,GHz
\citep{PCCS2_2016}. We checked that using APERFLUX, the recommended
photometry above 217\,GHz, the results at 353\,GHz do not change significantly.

For each bin we compared the distribution of polarized signals with that
of control  fields using the one-sided KS statistics. Generally, no signal was
detected for the faintest bins. When a signal was detected, the mean
polarized flux density was computed using Eq.~(\ref{eq:pol_flux}). This shows
another advantage of the analysis by flux density bins over the stacking
approach applied to the whole sample: it allows us to include for the final
estimates only the flux density bins that include signals; fainter objects whose
polarized emission is completely swamped by noise and fluctuations of other
components can be excluded from the analysis.

While the DETFLUX photomery was used for the binning, for estimating the
total  flux densities used to compute $\Pi_{\rm IDA}=(P/S)_{\rm median}$ we
adopted for uniformity the same approach as for the polarization signals: we summed signals in pixels within a radius of FWHM/2 from the source
position.

\begin{table}
\caption{Comparison of the median polarization degrees, $\Pi_{\rm IDA}=(P/S)_{\rm median}$,
at 44\,GHz yielded by the IDA method with those measured by \citet{Jackson2010} at 43\,GHz,
within the flux density range $S_{\rm range}$. The errors on $\Pi_{\rm IDA}$ correspond to
the 16th and 84th percentiles of the polarization degree distribution divided by
$N_{\rm bin}^{1/2}$, $N_{\rm bin}$ being the number of sources in the bin, $N_{\rm PCCS2}$
is the number of them with polarization measurements in the PCCS2, $D$ is the KS statistics.
The last column gives the probability of the null hypothesis (the distributions of signals of
sources and control fields are drawn from the same parent distribution) given by the one-sided KS test.
The bottom line refers to the full sample.}
\label{tab:jackinpccs2}
\begin{adjustbox}{width=0.49\textwidth}
\begin{tabular}{ccccccc}
\hline
$S_{\rm range}$ (mJy) &  $N_{\rm bin}$ & $N_{\rm PCCS2}$  & $\Pi_{\rm Jackson}$ & $\Pi_{\rm IDA}$  & $D$
  & Probability  \\
\hline
  695--958   &  \phantom{1}30 &  0   &  0.031 &  0.043 (+0.013, -0.008) & 0.400 &  $6.8\times 10^{-5}$ \\
  958--1290  &  \phantom{1}30 &  0   &  0.027 &  0.053 (+0.013, -0.010) & 0.433 &  $1.3\times 10^{-5}$ \\
 1290--1740  &  \phantom{1}30 &  0   &  0.031 &  0.021 (+0.009, -0.004) & 0.467 &  $2.1\times 10^{-6}$ \\
 1740--3510  &  \phantom{1}30 &  1   &  0.027 &   ---                    &   --- &  --- \\
 3510--23260 &  \phantom{3}15 &  4   &  0.031 &  0.033 (+0.002, -0.009) & 0.496 &  $6.2\times 10^{-4}$ \\
 \ \\
 695--23260  &       135      &  5   &  0.028 &  0.023 (+0.006, -0.002) & 0.437 &  $5.2\times 10^{-23}$ \\
\hline
\end{tabular}
\end{adjustbox}
\end{table}

\begin{table}
\caption{Comparison of the median polarization degrees, $\Pi$, at 100\,GHz yielded by
the IDA method with those measured by \citet{Agudo2014} at 86\,GHz. The columns have the
same meaning as in Table~\ref{tab:jackinpccs2}. The bottom line refers to the full sample.}
\label{tab:agudoinpccs2}
\begin{adjustbox}{width=0.49\textwidth}
\begin{tabular}{ccccccc}
\hline
$S_{\rm range}$ (mJy) &  $N_{\rm bin}$ & $N_{\rm PCCS2}$  & $\Pi_{\rm Agudo}$ & $\Pi_{\rm IDA}$  & $D$     & Probability  \\
\hline
   300--550  & \phantom{1}30  &         0    & 0.035   &     0.040 (+0.013, -0.007) &   0.400 &  $6.8\times 10^{-5}$ \\
   550--800  & \phantom{1}30  &         0    & 0.028   &     0.023 (+0.018, -0.004) &   0.400 &  $6.8\times 10^{-5}$ \\
   800-1090  & \phantom{1}30  &         1    & 0.028   &     0.028 (+0.012, -0.005) &   0.400 &  $6.8\times 10^{-5}$ \\
 1090--1620  & \phantom{1}30  &         2    & 0.032   &     0.009 (+0.010, -0.002) &   0.467 &  $2.1\times 10^{-6}$ \\
 1620--4850  & \phantom{1}30  &         5    & 0.026   &     0.030 (+0.005, -0.005) &   0.436 &  $1.1\times 10^{-5}$ \\
 4850--15710 & \phantom{13}6  &         5    & 0.036   &     0.034 (+0.015, -0.014) &   0.827 &  $2.7\times 10^{-4}$ \\
 \ \\
 300--15710  &         156    &        13    & 0.030   &     0.028 (+0.005, -0.002) &   0.359 &  $3.7\times 10^{-18}$ \\
\hline
\end{tabular}
\end{adjustbox}
\end{table}

\section{Validation with external data}
\label{sect:validation}

To validate our approach, we have exploited the ground-based polarization
measurements  by \citet{Jackson2010} and by \citet{Agudo2014} at frequencies
close to those of a \textit{Planck} channel.

At 43\,GHz \citet{Jackson2010} detected 167 sources. We divided those
detected by  \textit{Planck} in total flux density at 44\,GHz
\citep{PCCS2_2016} within the area specified above (135 sources) into bins
containing 30 sources each,  except for the brightest bin, which contains 15
sources. The KS test detected signal in \textit{Planck} polarization maps for
all bins, except for the 1740--3510\,mJy bin. All detections are highly
significant, as shown by the last column of Table~\ref{tab:jackinpccs2}. In
this table (and in the following tables), $D$ is the KS statistics, that is, the
largest discrepancy between the cumulative distributions of sources in a bin
and of control fields. The probability of the null hypothesis (no difference
between the two) was computed by approximating the distribution of
\begin{equation}
X^2=4\,D^2\,\frac{mn}{m+n}
\label{eq:Siegel}
\end{equation}
\noindent with the chi-square distribution with two degrees of freedom
\citep{SiegelCastellan1988}.  Here $m$ and $n$ are the numbers of sources and
of control fields, respectively.

For the full sample, the IDA method yielded estimates of the median
polarization  degrees consistent with those measured by \citet{Jackson2010},
within the errors that correspond to the 16th and 84th percentiles of the
polarization  degree distribution, divided by the square root of the number of
sources in the bin (cf. Table~\ref{tab:jackinpccs2}).\footnote{Note that
differences larger than statistical errors can be expected for measurements
made at different epochs, as a consequence of variability.}

\citet{Agudo2014} reported $\ge 3\,\sigma$ linear polarization detections of
183 sources at  86\,GHz (88\% of the sample detected in total flux density at
this frequency) and of 23 sources at 229\,GHz (13\% of those detected in total
flux density); 156 of the sources detected in polarization at 86\,GHz are
listed in the PCCS2 100\,GHz catalog within the area considered here. These sources were again divided into flux density bins containing 30 sources each,
except for the brightest bin, which contains 6 sources. Highly significant
polarization signals are detected for all bins (cf.
Table~\ref{tab:agudoinpccs2}). The method yields values of
the median polarization degrees in close agreement with the ground-based
measurements in this case as well.

Of the 23 sources detected in polarization at 229\,GHz by \citet{Agudo2014}, 18
are within  the region we considered, but one of them was not detected by
\textit{Planck} at 217\,GHz. For the remaining 17 sources, the KS test gives a
detection probability of 92\% (i.e., a probability of the null hypothesis,
sources extracted from the same distribution as control fields, of 8\%). The
median polarization degree measured on \textit{Planck} maps is $4.8 (+1.7,
-2.7)\%$ to be compared with $7.9 (+2.2, -1.8)\%$ measured by \citet{Agudo2014}
for the same sources. Given that only a small fraction of observed sources were
detected, it is quite likely that the detected sources, which are strongly
variable, were preferentially caught by \citet{Agudo2014} in a high-polarization phase. Hence it is not surprising that the median value derived
from \textit{Planck} maps, which are averages over five scans of the sky spanning
2.5 years, is somewhat lower.

\begin{figure*}
\centering
\includegraphics[width=\textwidth,clip]{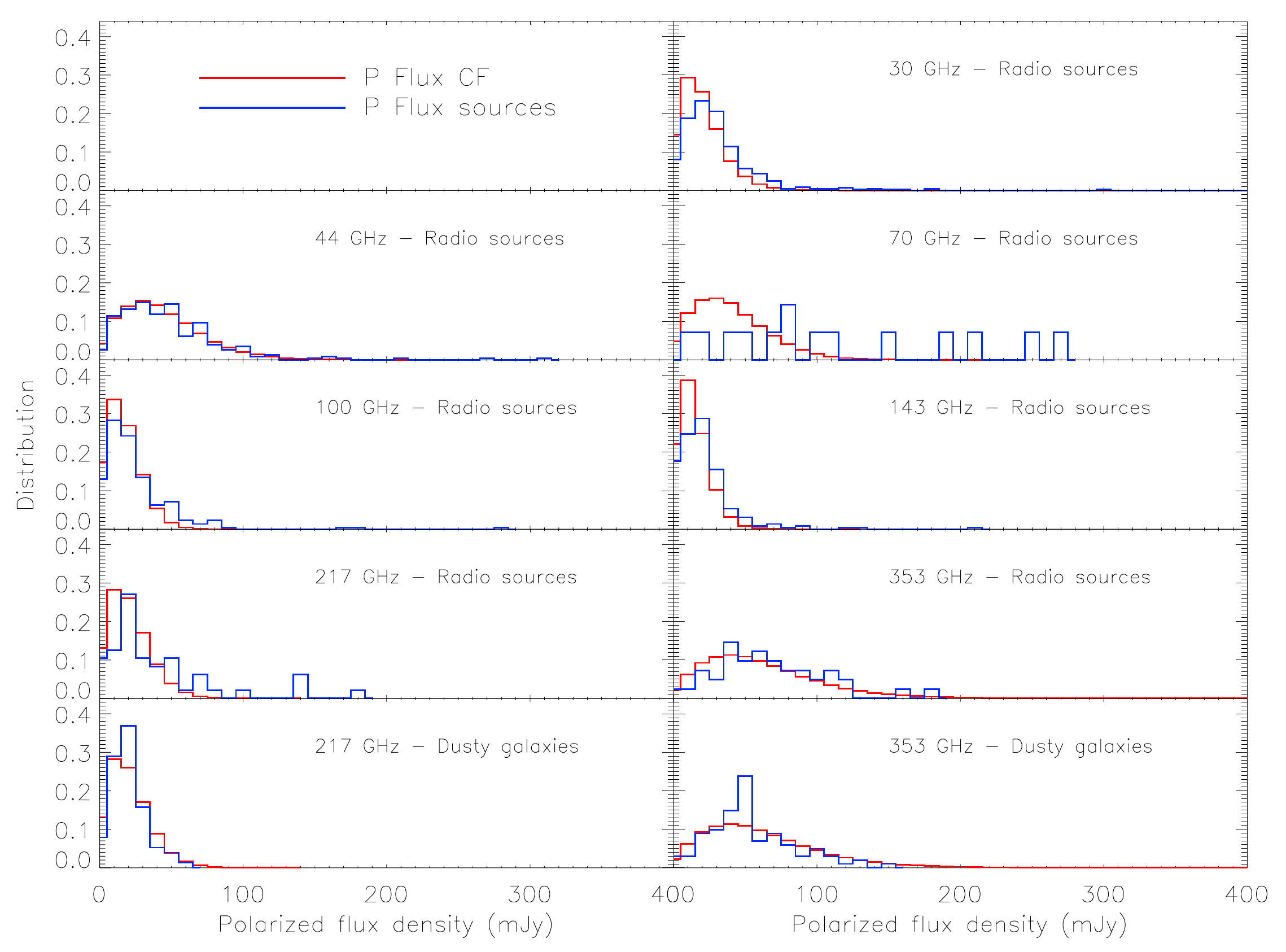}
\caption{Distribution of the polarized signals of PCCS2 radio sources and dusty galaxies
compared with those of control fields (signals within a radius of FWHM/2 from the corresponding centers).
The total flux density limits for radio sources are those listed in Table~\ref{tab:avg_vs_median}.
For dusty galaxies at 353\,GHz we have kept the same limit as for radio sources (784\,mJy).
At 217\,GHz, $S_{\rm lim, dusty}=300\,$mJy (see text).}
\label{fig:histo}
\end{figure*}

\begin{table}
\caption{Median polarization degrees, $\Pi_{\rm IDA}=(P/S)_{\rm median}$, at 30 GHz of PCCS2
radio sources in the region specified in Sect.~\ref{sect:method}. The columns have the same
meaning as in Table~\ref{tab:jackinpccs2}. The first bin is the faintest for which the test
detected a significant signal.}
\begin{adjustbox}{width=0.49\textwidth}
\begin{tabular}{cccccc}
\hline
$S_{\rm range}$ (mJy) &  $N_{\rm bin}$ & $N_{\rm PCCS2}$  & $\Pi_{\rm IDA}$  & $D$     & Probability  \\
\hline
596--621      &  29           &   0        &  0.054 (+0.017, -0.025)  &  0.345 & $1.0\times 10^{-3}$ \\
621--650      &  30           &   0        &  0.059 (+0.009, -0.015)  & 0.303 & $4.1\times 10^{-3}$ \\
650--682      &  30           &   1        &  0.038 (+0.014, -0.027)  & 0.400 & $6.8\times 10^{-5}$ \\
682--712      &  30           &   0        &  0.010 (+0.024, -0.010)  & 0.500 & $3.1\times 10^{-7}$ \\
712--764      &  30           &   1        &  0.036 (+0.021, -0.012)  & 0.300 & $4.5\times 10^{-3}$ \\
764--830      &  30           &   0        &  0.044 (+0.012, -0.019)  & 0.333 & $1.3\times 10^{-3}$ \\
830--894      &  30           &   0        &  0.042 (+0.010, -0.030)  & 0.400 & $6.8\times 10^{-5}$ \\
894--980      &  29           &   0        &  0.026 (+0.009, -0.015)  & 0.379 & $2.4\times 10^{-4}$ \\
980--1075     &  30           &   1        &  0.019 (+0.011, -0.019)  & 0.433 & $1.3\times 10^{-5}$ \\
1075--1210      &  30           &   0        &  0.036 (+0.007, -0.015)  & 0.333 & $1.3\times 10^{-3}$ \\
1210--1370      &  28           &   2        &  0.031 (+0.005, -0.013)  & 0.321 & $3.1\times 10^{-3}$ \\
1370--1636      &  30           &   5        &  0.032 (+0.007, -0.006)  & 0.371 & $2.6\times 10^{-4}$ \\
1635--2125      &  30           &   3        &  0.025 (+0.004, -0.007)  & 0.346 & $7.7\times 10^{-4}$ \\
2125--3800      &  30           &   6        &  0.026 (+0.004, -0.005)  & 0.440 & $9.2\times 10^{-6}$ \\
3800--30370      &  22           &  16        &  0.031 (+0.005, -0.005)  & 0.750 & $1.8\times 10^{-11}$ \\
\hline
\end{tabular}
\end{adjustbox}
\label{tab:pccs2_030}
\end{table}

\begin{table}
\caption{Same as in Table~\ref{tab:pccs2_030}, but at 44 GHz.}
\label{tab:pccs2_044}
\begin{adjustbox}{width=0.49\textwidth}
\begin{tabular}{cccccc}
\hline
$S_{\rm range}$ (mJy) &  $N_{\rm bin}$ & $N_{\rm PCCS2}$  & $\Pi_{\rm IDA}$  & $D$     & Probability  \\
\hline
873--945    &     30  &   0  &    0.035 (+0.017, -0.025) &  0.400  & $6.8\times 10^{-5}$  \\
945--1042   &     30  &   0  &    0.038 (+0.016, -0.038) &  0.467  &  $2.1\times 10^{-6}$ \\
1042--1162  &     30  &   0  &    0.024 (+0.013, -0.024) &  0.433  &  $1.3\times 10^{-5}$ \\
1162--1320  &     30  &   0  &    0.045 (+0.019, -0.024) &  0.367  &  $3.2\times 10^{-4}$ \\
1320--1645  &     30  &   0  &    0.021 (+0.008, -0.004) &  0.467  &  $2.1\times 10^{-6}$ \\
1645--2100  &     30  &   1  &        ---                &  ---    &   ---                \\
2100--3700  &     30  &   2  &    0.003 (+0.008, -0.003) &  0.500  &  $3.1\times 10^{-7}$ \\
3700--29204 &     18  &   6  &    0.018 (+0.007, -0.004) &  0.508  &  $9.4\times 10^{-5}$ \\
\hline
\end{tabular}
\end{adjustbox}
\end{table}

\begin{table}
\caption{Same as in Table~\ref{tab:pccs2_030}, but at 100 GHz.}
\label{tab:pccs2_100}
\begin{adjustbox}{width=0.49\textwidth}
\begin{tabular}{cccccc}
\hline
$S_{\rm range}$ (mJy) &  $N_{\rm bin}$ & $N_{\rm PCCS2}$  & $\Pi_{\rm IDA}$  & $D$     & Probability  \\
\hline
663--728    &  30    &    0   &  0.037 (+0.016, -0.037)   & 0.433 &  $1.3\times 10^{-5}$ \\
728--812    &  30    &    0   &  0.023 (+0.015, -0.016)   & 0.400 &  $6.8\times 10^{-5}$ \\
812--913    &  30    &    0   &  0.010 (+0.009, -0.010)   & 0.500 &  $3.1\times 10^{-7}$ \\
913--1030   &  30    &    1   &  0.008 (+0.016, -0.008)   & 0.500 &  $3.1\times 10^{-7}$ \\
1030--1300  &  30    &    1   &    ---                    &  ---  &      ---             \\
1300--1619  &  30    &    3   &  0.027 (+0.012, -0.019)   & 0.400 &  $6.8\times 10^{-5}$ \\
1619--2850  &  30    &    2   &  0.028 (+0.005, -0.006)   & 0.396 &  $8.4\times 10^{-5}$ \\
2850--7714  &  14    &    7   &  0.026 (+0.005, -0.006)   & 0.621 &  $2.1\times 10^{-5}$ \\
\hline
\end{tabular}
\end{adjustbox}
\end{table}

\begin{table}
\caption{Same as in Table~\ref{tab:pccs2_030}, but at 143 GHz.}
\label{tab:pccs2_143}
\begin{adjustbox}{width=0.49\textwidth}
\begin{tabular}{cccccc}
\hline
$S_{\rm range}$ (mJy) &  $N_{\rm bin}$ & $N_{\rm PCCS2}$  & $\Pi_{\rm IDA}$  & $D$     & Probability  \\
\hline
525--580     &   30  &   0   & 0.005 (+0.021, -0.005)  &  0.500 & $3.1\times 10^{-7}$ \\
580--648     &   30  &   0   & 0.029 (+0.012, -0.012)  &  0.333 & $1.3\times 10^{-3}$ \\
648--750     &   30  &   0   & 0.002 (+0.012, -0.002)  &  0.500 & $3.1\times 10^{-7}$ \\
750--820     &   29  &   1   & 0.040 (+0.005, -0.024)  &  0.379 & $2.4\times 10^{-4}$ \\
820--1048    &   27  &   2   & 0.028 (+0.012, -0.016)  &  0.370 & $6.1\times 10^{-4}$ \\
1048--1290   &   30  &   1   & 0.043 (+0.007, -0.009)  &  0.430 & $1.5\times 10^{-5}$ \\
1290--1650   &   30  &   5   & 0.022 (+0.006, -0.006)  &  0.294 & $5.5\times 10^{-3}$ \\
1650--5305   &   21  &   6   & 0.030 (+0.003, -0.009)  &  0.412 & $7.9\times 10^{-4}$ \\
\hline
\end{tabular}
\end{adjustbox}
\end{table}


\begin{table}
\caption{Median polarization degrees of radio sources at the \textit{Planck} polarization-sensitive frequencies.
All values were corrected for the small biases discussed in Sect.~\ref{sect:sims}. }
\label{tab:avg_vs_median}
\begin{adjustbox}{width=0.49\textwidth}
\begin{tabular}{ccccccc}
\hline
$\nu$ (GHz) & $S_{\rm min}$ (mJy)  &  $N$ & $N_{\rm PCCS2}$  & $\Pi_{\rm IDA}$  & $D$     & Probability  \\
\hline
\phantom{1}30        & \phantom{1}596   &    438    &     35   &  0.033 (+0.003, -0.003) &   0.311 & $8.4\times 10^{-37}$  \\
\phantom{1}44        & \phantom{1}873   &    228    & \phantom{1}9   &  0.022 (+0.006, -0.011) &   0.447 & $5.1\times 10^{-40}$ \\
\phantom{1}70        & 3749  &  \phantom{1}15 &   \phantom{1}4   &  0.028 (+0.006, -0.006) &   0.618 & $1.0\times 10^{-5}$  \\
 100           & \phantom{1}663   &   224    &     14   &  0.019 (+0.005, -0.005) &   0.406 & $9.0\times 10^{-33}$ \\
 143        & \phantom{1}525   &      227    &     15   &  0.029 (+0.003, -0.005) &   0.348 & $1.5\times 10^{-24}$ \\
 217        & 1085  &    \phantom{1}48    &     \phantom{1}8   &  0.031 (+0.004, -0.008) &   0.292 & $2.8\times 10^{-4}$  \\
 353        & 784   &    \phantom{1}41    &  \phantom{1}1      &  0.030  (+0.016, -0.020) &  0.415 &  $7.6\times 10^{-7}$ \\
\hline
\end{tabular}
\end{adjustbox}
\end{table}

\section{Results}\label{sect:results}

\subsection{Radio sources}

We detected significant polarization signals of PCCS2 radio sources in the
region specified in  Sect.~\ref{sect:method} for more than two flux density
bins at 30, 44, 100, and 143\,GHz. The median polarization degrees yielded by
the IDA method for each bin are reported in
Tables~\ref{tab:pccs2_030}--\ref{tab:pccs2_143}.  $\Pi_{\rm IDA}$ was computed
using the median of $\Pi$, Eq.~(\ref{eq:poldeg}), over the considered sample.
The errors correspond to the 16th and 84th percentiles of the polarization
degree distribution, divided by the square root of the number of sources in the
bin.

The first bin in each table is the faintest for which we detected a
significant  \textit{\textup{positive}} signal. The significance of the detections (not
to be confused with the S/N) is at least at the $\simeq
3\,\sigma$ level and in most cases is much higher, as shown in the last column
of Tables~\ref{tab:pccs2_030}--\ref{tab:pccs2_143}. There is no significant
dependence of $\Pi_{\rm IDA}$ on flux density at any frequency.

In Fig.~\ref{fig:histo} the distributions of polarized signals of radio sources
is  compared with those of control fields. The source distributions refer to
the total flux densities above the $S_{\rm lim}$ listed in
Table~\ref{tab:avg_vs_median}. The shift of the distributions toward higher
values of the polarized flux density, compared to control fields, can be
perceived by eye.

\begin{figure}
\centering
\includegraphics[width=0.5\textwidth,clip]{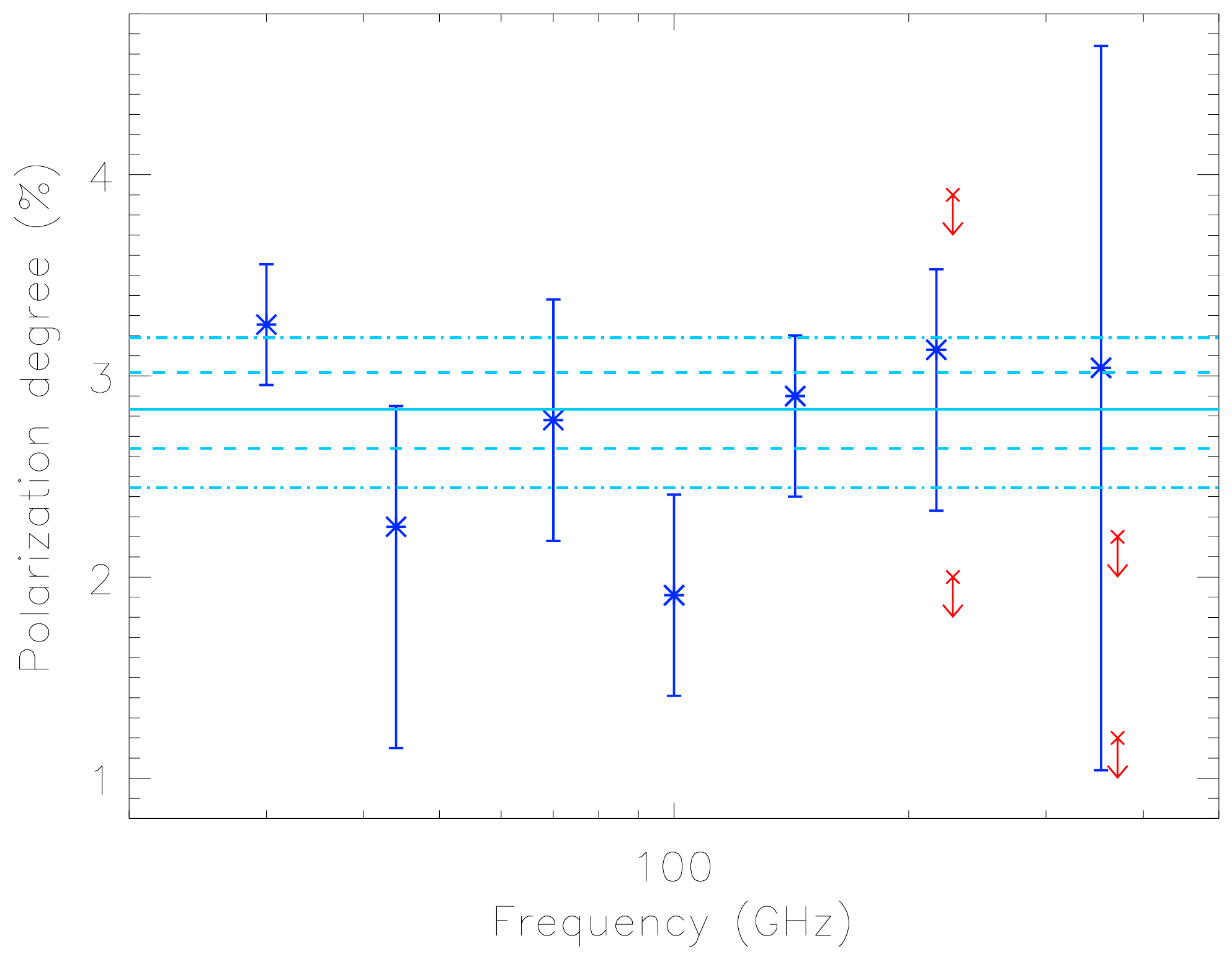}
\caption{Median polarization degrees for radio sources at \textit{Planck} frequencies with their errors
(blue data points), and upper limits (90\% and 68\% confidence) for dusty galaxies (red arrows).
All values were corrected for the small biases discussed in Sect.~\ref{sect:sims}.
The solid blue line shows the minimum $\chi^2$ value of the median polarization degree of radio sources,
assumed to be frequency independent. The dashed and dot-dashed lines show the 68\% and 95\% limits, respectively.
 }
\label{fig:poldeg}
\end{figure}

Table~\ref{tab:avg_vs_median} and Fig.~\ref{fig:poldeg} show the median
polarization  degrees, with their errors, for the full flux density ranges over
which a signal was detected, at all the \textit{Planck} polarization-sensitive
frequencies. The KS test always detects signals with high significance, and the
estimates have an S/N $>3$ except at 44 and 353\,GHz.  Assuming
that $\Pi_{\rm IDA}$ is frequency independent, a minimum $\chi^2$ fit gives
$\langle \Pi_{\rm IDA}\rangle = 2.83 (+0.18, -0.19)\%$ with a reduced $\chi^2$, $\chi_r^2$
$\simeq$ 1.08.\footnote{When the bias correction was not applied, we obtained
$\langle \Pi_{\rm IDA}\rangle = 2.75\%$ and a slightly larger $\chi_r^2$ ($\simeq$ 1.2).}

The median polarization degree does not show any statistically significant
frequency  dependence, consistent with the results of \citet{Battye2011} and
\citet{Galluzzi2017a, Galluzzi2018}, but not with the increase between 86 and
229\,GHz by factors of $\sim 1.6$ or $\sim 2.6$ claimed by \citet{Agudo2014}
and by \citet{2018MNRAS.473.1850A}, respectively.

Our results are generally in good agreement with ground-based high sensitivity
polarization measurements at nearby frequencies. At 33\,GHz,
\citet{Galluzzi2018} found for their fainter sample $\langle
\Pi\rangle=1.85\%$ with first and third quartile values of 1.17\% and 3.29\%,
respectively, to be compared with our median values at 30 GHz of $3.3\%$. At
43\,GHz, \citet{Battye2011} found $\Pi_{\rm median}=2.25\%$; our median value at
44\,GHz is $2.2\%$. At 86 GHz, \citet{Agudo2014} found a median polarization
degree, including upper limits, $\Pi_{\rm median}=2.9\%$ with first and third
quartiles of 1.8\% and 4.8\%, respectively, to be compared with our median
value at 100\,GHz of $1.9\%$.

We also agree with \citet{Bonavera2017radio}, who found that the average
fractional  polarization of radio sources is approximately
frequency independent in the \textit{Planck} range and derived a weighted
\textit{\textup{mean}} over all the channels of $\Pi=3.08\%$. Their \textup{median}
values, $\Pi=1.9\%,$ are slightly lower, and close to ours.

\subsection{Dusty galaxies}

The same approach was applied to investigate the polarization properties of
dusty galaxies.  Samples of these objects within the same area considered for
radio sources were extracted from the PCCS2 at 217 and 353\,GHz. The selection
of dusty galaxies was made taking into account only sources without a
counterpart at 143\,GHz, where radio sources dominate, as mentioned above.

We collected 616 and 678 dusty galaxies brighter than 152\,mJy at 217\,GHz and
than 304\,mJy at 353\,GHz. As in the case of radio sources, these objects were
subdivided into total flux density bins containing 30 dusty galaxies
each, except for
the brightest bins, which contain 16 and 18 objects, respectively.

The KS test did not detect any significant positive shift of the distribution
of  polarized flux densities compared to that of control fields. Examples are
shown in the bottom panels of Fig.~\ref{fig:histo}. At 353\,GHz, we adopted $S_{\rm lim}=784\,$mJy, the same as for radio sources. The 68\% and
90\% confidence upper limits to the polarization degree are $\Pi_{353, \rm
dusty}\simlt 1.1\%$ and $\simlt 2.2\%$, respectively. We  checked that
these upper limits are stable by varying $S_{\rm lim}$.\footnote{In particular, we
found  very stable 68\% (90\%) upper limits for $S_{\rm lim}$ between
$300\,$mJy ($500\,$mJy) and $1200\,$mJy.}

At 217\,GHz, there are no objects above 1085\,mJy (the limit for radio sources)
and only 11 objects above 500\,mJy. Thus, the histogram refers to $S_{\rm lim,
dusty}=300\,$mJy. Correcting for the small bias, the 68\% and 90\% confidence upper limits are
$\Pi_{217, \rm dusty}\simlt 2.0\%$ and $\simlt 3.9\%$, respectively.
The looser limits at 217\,GHz, compared to those at 353\,GHz, reflect the weakening of dusty galaxy
flux densities. These limits do not vary appreciably when we decrease $S_{\rm
lim}$. The number of sources rapidly decreases for higher $S_{\rm lim}$, and
the limits become unstable.

We therefore do not confirm the conclusion by \citet{Bonavera2017dusty}, who
derived \textit{\textup{mean}} fractional polarizations of $(3.65 \pm 0.66)\%$ and
$(3.10 \pm 0.75)\%$, and corresponding \textit{\textup{median}} values of $(2.0 \pm
0.8)\%$ and $(1.3 \pm 0.7)\%$, at 353\,GHz and at 217\,GHz, respectively. We note,
however, that their median values are consistent with our 68\% confidence upper
limit at 217\,GHz and with our 90\% confidence upper limit at 353\,GHz.

\begin{figure*}
\centering
\includegraphics[width=0.9\textwidth,clip]{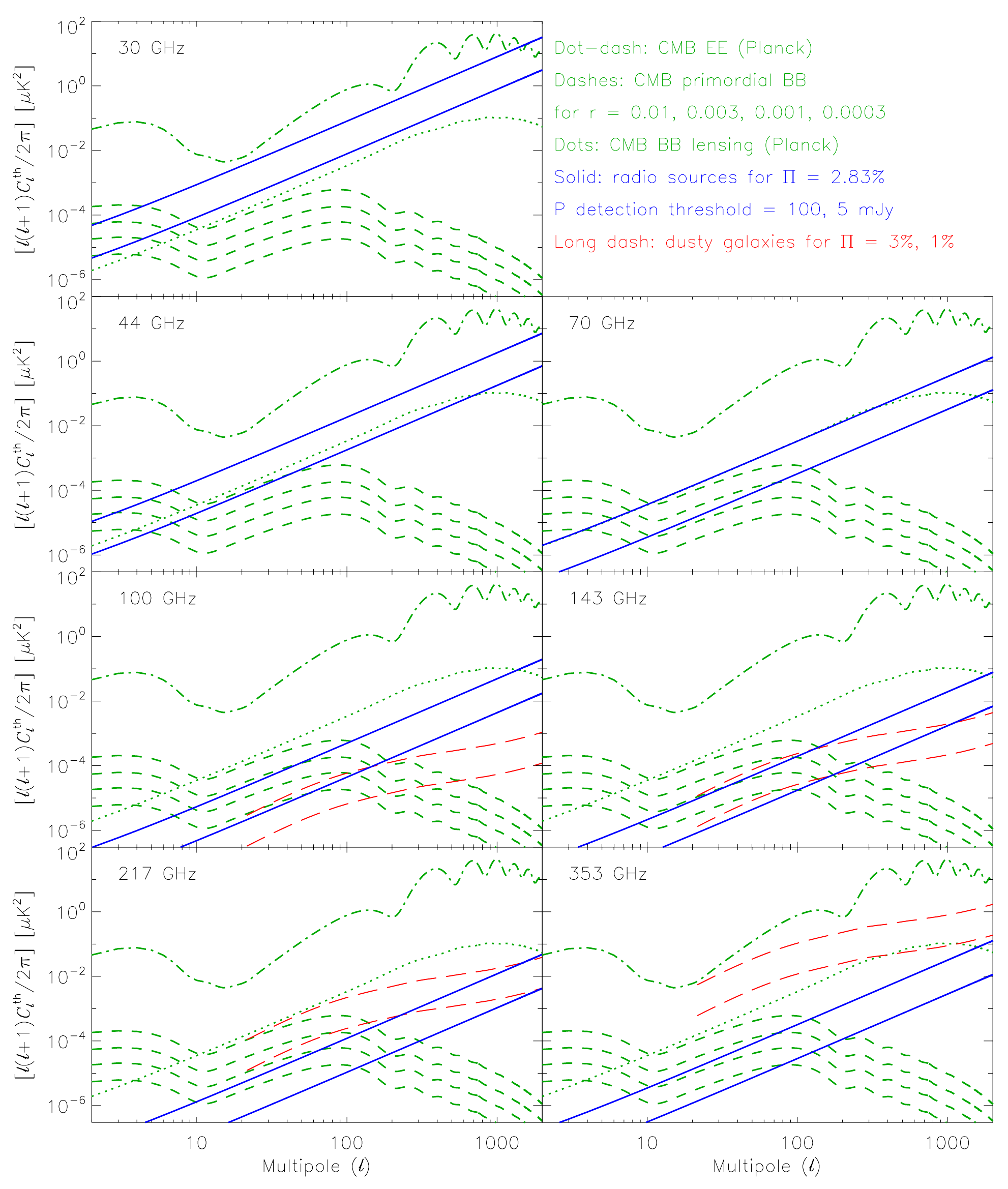}
\caption{Polarization E/B-mode power spectra of radio sources and of dusty galaxies
compared with the CMB E mode and B mode for four values of the tensor to scalar ratio
$r = 0.01, 0.003, 0.001, 0.0003$ (from top to bottom); see text and legend for details.
Since unresolved point sources on average contribute equally to E and B mode, their
total power spectra have been divided by a factor of 2.}
\label{fig:power_spectra}
\end{figure*}

\subsection{Validation through simulations}
\label{sect:sims}

Although the comparison with the polarization measurements by
\citet{Jackson2010} and by \citet{Agudo2014} has shown that the IDA approach
accounts for and remarkably well corrects the noise bias (cf.
Sect.~\ref{sect:validation}), Eq.~(\ref{eq:pol_flux}) can still be affected by
bias residuals, whose magnitude can be estimated by means of Monte Carlo simulations.

In the ideal case, the polarized flux density of the $i$--th source is
$P_{i}^{2} = Q_{i}^{2} + U_{i}^{2}$  and, for an ensemble of $N_s$ sources, the
median polarization degree is
\begin{equation}
\Pi_{\rm  median} = \left. \frac{P_{i}}{S_{i}} \right |_{\rm median}.
\label{eq:pibiascorr}
\end{equation}
\noindent In our approach, Eq. (\ref{eq:pibiascorr}) becomes
\begin{equation}
\Pi = \left. \frac{(P_{{\rm obs}, i}^{2} - P_{\rm{{\rm CF}, median}}^{2})^{1/2}}{S_{{\rm obs},i} - S_{\rm{{\rm CF}, median}}} \right |_{\rm median},
\label{eq:pibias}
\end{equation}
\noindent where $(P\, {\rm or}\, S)_{\rm{{\rm CF}, median}}$ is the
median over all the  $N_{\rm CF}$ control fields (with $N_{\rm CF}>>N_{s}$).
The signal in the $j$--th control field depends on noise and ``foregrounds'' 
(here including the CMB as well), that is, 
$P_{{\rm CF}, j}^{2} = (Q_{n} + Q_{f})_{j}^{2} + (U_{n} + U_{f})_{j}^{2}$ and $S_{{\rm CF}, j} = (S_{n} +
S_{f})_{j}$, while for the $i$-th source, we have $P_{{\rm obs}, i}^{2} = (Q_{s}
+ Q_{n} + Q_{f})_{i}^{2} + (U_{s} + U_{n} + U_{f})_{i}^{2}$ and $S_{{\rm obs},
i} = (S_{s} + S_{n} + S_{f})_{i}$. The ($Q$, $U$) cross-product terms
of source with noise and foregrounds are not subtracted in the numerator of
Eq.~(\ref{eq:pibias}) and their contribution is not fully negligible, in
principle, even though they are uncorrelated, hence vanishing on average, and
expected to be significantly suppressed in the median.

To quantify these potential residuals, we performed Monte Carlo simulations
consisting of 1000 realizations. Since diffuse foregrounds vary especially with
Galactic latitude, we analyzed the following cases: Galactic latitude $|b|
> 20^{\circ}$, $|b| \in (20, 40)^{\circ}$, $|b|\in (40, 60)^{\circ}$ , and
$|b| > 60^{\circ}$. The simulated sources in the mock catalogs
(chosen in number as in column 3 of Table \ref{tab:avg_vs_median})
were randomly located at the positions of some control fields,
in order to avoid regions where real sources are detected. We generated source flux densities in the same
ranges as real sources at each frequency, in keeping with their differential
number counts.

Mock radio sources and dusty galaxies were assigned $\Pi_{\rm RS} =
2.75\%$ and $\Pi_{\rm dusty} = 1 \%$, respectively; ($Q_{i}$, $U_{i}$) were
derived from a uniform distribution of polarization angles. Applying the IDA
method, we generally found for radio sources bias values of $\delta \Pi_{\rm
RS} \simeq -(0.05-0.06)\%$ except at 44\,GHz (where \textit{Planck} maps are
particularly noisy) and at 353\,GHz (where the statistics is poor); at these
frequencies we have $\delta \Pi_{\rm RS, 44} \simeq -0.12 \%$ and $\delta
\Pi_{\rm RS, 353} \simeq -0.31\%$. For dusty galaxies, simulated only at 217
and 353\,GHz, we found $\delta \Pi_{\rm dusty, 217} \simeq -0.1\%$ and a
negligible value at 353\,GHz.

These results refer to $|b| > 20^{\circ}$, but the bias amplitudes
do not show any significant dependence on Galactic latitudes, except at
353\,GHz, where they somewhat increase for $|b|\in (40, 60)^{\circ}$ and $|b|
> 60^{\circ}$ because of the poorer statistics of control fields that is due to the smaller sky fraction. In general, the bias
values are much lower than the uncertainties on $\Pi$; nevertheless, the
median polarization degrees in Table~\ref{tab:avg_vs_median} and in
Fig.~\ref{fig:poldeg} include this correction. The errors on the retrieved
median values are about a factor of 2 lower than those estimated for real
sources. The reason probably is that the simulations do not allow for
the intrinsic dispersion of the source fractional polarization.

\subsection{Power spectra}

Figure~\ref{fig:power_spectra} compares the polarization power spectra of radio
sources  (solid blue lines) and of dusty galaxies (dashed red lines, shown only
for $\nu \ge 100\,$GHz) with the CMB power spectra. The dot-dashed and dotted
green lines show the E-mode and lensing B-mode power spectra for the
\textit{Planck} best-fit cosmological parameters
\citep{PlanckCollaborationParameters2016}. The dashed green lines show the
primordial B-mode power spectra for four values of the tensor to scalar ratio.

The power spectra of sources are the sum of Poisson and clustering
contributions.  Poisson fluctuations have a white-noise power spectrum,
independent of the multipole number $\ell$:
\begin{equation}
C_{\ell,\rm Poisson}={\mathlarger\int}_{\!\!\!0}^{S_d} {dN\over dS}\, S^2\, dS \,,
\label{eq:Poisson}
\end{equation}
\noindent where $dN(S)/dS$ are the differential number counts per steradian of
sources weaker  than the detection limit $S_d$.

In the case of extragalactic radio sources, the contribution of clustering can
be neglected  \citep[cf., e.g., ][]{Delabrouille2013}. To compute the Poisson
power spectrum, counts in polarized flux density were estimated from
counts in total flux density, $S$,  adopting a polarization fraction of
2.83\%,
that is, setting $S_p=0.0283\,S$.
For the counts in total flux density, we
exploited the models by \citet{DeZotti2005} up to 70\,GHz and by
\citet{Tucci2011} at higher frequencies. In the relevant flux-density range, the
slope of the $dN(S)/dS$ of extragalactic radio sources is $\simlt 2$, so that
the largest contribution to $C_{\ell,\rm Poisson}$ comes from flux densities
just below the detection limits in polarization. Figure~\ref{fig:power_spectra}
shows two cases, considering the polarized flux density detection limits
expected for next-generation CMB experiments \citep[cf. Fig.~8 of
][]{DeZotti2018}.

Conversely, in the case of dusty galaxies, the clustering contribution
dominates most of the multipole range of interest here. The power spectra
in polarized flux density were derived from those in total flux density
given by the \citet{Cai2013} model that accurately reproduce both the
\textit{Planck} \citep{PlanckCollaboration2011CIB, PlanckCollaborationXXX2014}
and the \textit{Herschel} \citep{Viero2013} measurements. For these objects, the
counts are very steep (slope $>3$), so that the main contribution to the power
spectrum comes from faint sources, or in other words, the power spectrum amplitude is
essentially independent of $S_d$. Again, the power spectra in polarization were scaled from those in total flux density for two choices of the
polarization degree, 3\%, close to the mean value found by
\citet{Bonavera2017dusty}, and 1\%, close to our 68\% confidence upper limit at
353\,GHz. For the sake of illustration, we applied this approach down to
100\,GHz. We note, however, that already at 143\,GHz, the dusty galaxies in
\textit{Planck} maps are too faint to allow the derivation of reasonably
accurate constraints on their average polarization degree.

Unresolved point sources contribute, on average, equally to E- and B-mode power
spectra.  Thus, for comparison with the CMB polarization modes
(Fig.~\ref{fig:power_spectra}), the total power spectra discussed above were divided by a factor of 2.

The contamination of CMB polarization maps by extragalactic radio sources was
previously  discussed by \citet{Tucci2004}, \citet{TucciToffolatti2012}, and
\citet{Curto2013}. Our analysis agrees with their conclusion that these objects
are not a strong contaminant to the CMB E-mode polarization, but can constrain
the detection of cosmological B-modes if $r\simlt 0.01$.

Estimates of the power spectrum of dusty galaxies were presented by
\citet{Curto2013},  who assumed an average polarization level of 1\%, and by
\citet{Bonavera2017dusty}. The latter authors used log-normal distributions for
the polarization degrees and took into account only the Poisson contributions.
Extrapolating to lower frequencies the estimates or the upper limits obtained
at 217 and 353\,GHz, we find that the contamination by dusty galaxies may be
comparable to that of radio sources at 100--143\,GHz; it becomes dominant at
higher frequencies and rapidly fades away at lower frequencies.

\section{Conclusions}
\label{sect:conclusions}

We have revisited the estimates of the mean polarization fraction of
extragalactic  sources (radio-loud AGNs and dusty galaxies) based on data from
the \textit{Planck} polarization maps at 30, 44, 70, 100, 143, 217, and
353\,GHz. Although the earlier analyses by \citet{Bonavera2017radio,
Bonavera2017dusty} based on stacking techniques were carefully made, there are
several tricky aspects and subtleties that call for an independent analysis.
This is particularly important in relation to the forthcoming or proposed CMB
polarization experiments aimed at detecting primordial B modes.

The importance of a careful control of foregrounds has been demonstrated by
detailed sky simulations \citep{Remazeilles2018}, which included Galactic and
extragalactic polarized emissions in addition to the CMB, based on
state-of-the-art observations. These simulations have shown that for $r$ at
the per thousand level (i.e., at the level predicted by models currently of
special interest), or smaller (as in the case of large-field inflation with
super-Planckian field excursions), the overall uncertainty on this parameter
can be dominated by the contamination of unresolved polarized extragalactic
sources. An insufficiently accurate characterization of this component could
lead to a bias in the reconstruction of the primordial CMB B-mode signal.

Our independent reanalysis overcomes the two most delicate aspects of the
application of stacking techniques: the approximation of the average
polarization fraction, $\langle\Pi\rangle=\langle P/S \rangle$, with the ratio
of the mean polarized flux density to the mean total flux density,  $\langle P
\rangle/\langle S \rangle$, and the need of simulations to correct for the
noise bias.

Our approach considered the objects one by one. This allowed us to identify the
flux  density range that contributes significantly to the polarization signal;
thus we can exclude fainter objects from the analysis that may affect results
because of the serious limitation by noise and background signals. For objects
above the flux density threshold, we directly  computed the median $P/S$.  We find
that the method allows us to detect, on \textit{Planck} maps, mean polarized
flux densities at few tens of mJy levels. For comparison, the detection limits
in total intensity are at the few to several hundred mJy levels \citep[cf.
Table\,13 of][]{PCCS2_2016}.

In addition, the subtraction of the median of the polarization signal of control
fields largely corrects for the contributions of the noise and of the other
polarized components (CMB and Galactic emissions).
By means of simulations, we have found that residual biases on the
median polarization fractions are generally below 0.1\%, which
is much smaller than
the estimated errors.

For radio sources, we find a median polarization degree, averaged over
frequencies,  $\Pi_{\rm IDA, median}\simeq 2.83\%$, in good agreement with
\citet{Bonavera2017radio} as well as  with the ground-based measurements of a
fainter sample at 33 and 38\,GHz \citep{Galluzzi2018} and of a bright sample at
86\,GHz \citep{Agudo2014}. We do not find any significant dependence of $\Pi$
on either flux density or frequency, in agreement with earlier analyses at
frequencies up to 43\,GHz, but not in agreement with the increase of $\Pi$ from 86 to 229
GHz claimed by \citet{Agudo2014, 2018MNRAS.473.1850A}.

At variance with \citet{Bonavera2017dusty}, we do not detect any polarization
signal  from dusty galaxies, although their median values are consistent with
our upper limits. For these objects we derive a 90\% confidence upper limit at
353\,GHz $\Pi_{\rm dusty}\simlt 2.2\%$. The upper limit at the same confidence
level is looser ($\simlt 3.9\%$) at 217\,GHz, where dusty galaxies are
substantially fainter.

The contamination of CMB maps in polarization by extragalactic sources is
dominated  by radio-loud AGNs up to $\sim 100\,$GHz. The amplitude of their
power spectra depends on their detection limit in polarization, $S_d$. For the
values of $S_d$ expected for next-generation CMB experiments, we confirm that
at $\simeq 70\,$GHz, that is, in correspondence to the minimum Galactic emission,
the point source (radio source) contamination is well below primordial E modes,
as found by previous analyses. On the other hand, it is close to the level of
lensing B modes and of primordial B modes for $r\simeq 0.01$. The contribution
of dusty galaxies to the point source power spectra is still poorly
constrained, but may be substantial or even dominant at $\simgt 100\,$GHz.

\begin{acknowledgements}
Thanks are due to Z.-Y. Cai and to M. Tucci for having provided
the  power spectra of dusty galaxies and the high-frequency counts of radio
sources, respectively, yielded by their models. We also thank the anonymous referee
for comments that helped improve the paper. We gratefully acknowledge
financial support from ASI/INAF agreement n.~2014-024-R.1 for the {\it Planck}
LFI Activity of Phase E2, from the ASI/Physics Department of the university of
Roma--Tor Vergata agreement n. 2016-24-H.0 for study activities of the Italian
cosmology community and from the Italian Ministero dell'Istruzione,
Universit\`a e Ricerca through the grant `Progetti Premiali 2012-iALMA' (CUP
C52I13000140001). Some of the results in this paper have been derived using the
HEALPix \citep{Gorski2005} package.
\end{acknowledgements}

%
%
   \bibliographystyle{aa} 
   \bibliography{tiziana_ref} 

\end{document}